\documentclass[11pt]{article}
\usepackage{moriond,epsfig}
\usepackage[usenames]{color}

\def\be{\begin{equation}}
\def\ee{\end{equation}}
\def\bea{\begin{eqnarray}}
\def\eea{\end{eqnarray}}

\begin{document}
\vspace*{4cm}
\title{EFFECTIVE COUPLINGS APPROACH TO NEUTRALINO DARK MATTER RELIC DENSITY}

\author{Suchita Kulkarni}

\address{LPSC, 53 Av. des Martyrs, 38000, Grenoble, France}

\maketitle\abstracts{
In this work, we analyze the electroweak loop corrections to
the Neutralino dark matter relic density in the framework of
effective coupling. In the first part, we comment on the generic
features of the corrections and quantitative changes to the predicted
relic density. We analyze the correlation between the
characteristics of effective couplings to the nature of
neutralino. Effective couplings, however, absorb only the most dominant one loop
corrections and are not an exact calculation. In the second part, we assess the validity of effective couplings by comparing them to
the full one loop calculations in various regions of parameter space.
}

\section{Introduction}
Within the framework of $\Lambda$CDM, the existence of dark matter is well established. Although, the exact nature of this omnipresent dark matter is yet to be identified, its relic density is well measured and measurements from Planck satellite will reduce the uncertainly to $1\%$. This is good news because the abundance of dark matter not only depends on the astrophysical parameters like the Hubble constant and the velocity distribution, but also on the particle physics properties like it's mass and the cross-section. If dark matter is cold in nature then it's relic density $\Omega\,h^2$ is inversely proportional to the thermal average of the annihilation cross-section $\langle\sigma v \rangle$. 
%\begin{equation}
%\Omega\, h^2 \propto \frac{1}{\langle\sigma v \rangle}
%\end{equation}
Any changes in the theoretically predicted cross-section will thus change the relic density. In the light of upcoming results from Planck, a theoretically precise calculation of the relic density plays an important role in determining the favored regions of theory parameter space.

Various theories beyond the Standard Model of particle physics (SM) seek an explanation for dark matter. Supersymmetry is one of the most elegant extensions of the SM not only because it provides us with possible dark matter candidate but also because it gives us an explanation of the hierarchy problem in the higgs sector.~\cite{Drees:2004jm} Different ways of SUSY breaking at a high scale can lead to different dark matter candidates, the lightest neutralino or gravitino for example. Under the assumption of R-parity conserving SUSY, neutralino is the most widely studied dark matter candidate. However, the annihilation cross-section of neutralino is often taken at tree-level and changes due to loop-corrections can lead to a significant change in the predicted relic density. A full one loop analysis is numerically exhaustive due to large number of diagrams involved in annihilation cross-section while computing the relic density. 

In this work, we assess the impact of electroweak loop-corrections to neutralino dark matter via effective couplings and compare the validity of this approach to the full one loop analysis. 

\section{Effective couplings for neutralino}
We implement effective couplings for the neutralino-fermion-sfermion vertex, as first discussed by Guasch {\it et.al.}~\cite{Guasch:2002ez}  Analogous to the oblique corrections in the SM, they are flavor independent, finite corrections to the neutralino-fermion-sfermion vertex. 

The neutralino is a combination of supersymmetric counterparts of the gauge and Higgs boson, namely the gauginos and the higgsinos. After the $SU(2)_L\times U(1)_Y $ symmetry breaking, the bino, wino and higgsino mix to form four neutralinos, the lightest one being of interest to us as a dark matter candidate.  
%In the charge basis the neutralino mass matrix is given by: 
 %   \begin{equation} 
%	\mathcal{M} = \left( \begin{array}{cccc} M_1 & 0 & - \mz \sw c_{\beta} & \mz \sw s_{\beta} \\
%	    0 & M_2 & \mz \cw c_{\beta} & -\mz \cw s_{\beta} \\
%	    -\mz \sw c_{\beta} & \mz \cw c_{\beta} & 0 & -\mu\\
%	    \mz \sw s_{\beta} & -\mz \cw s_{\beta} & -\mu & 0 \\
%	    \end{array} \right) 
    % \end{equation}
    % with $M_1, M_2,\mu$ denoting the bino, wino and higgsino components of the neutralino. 
%This is diagonalized to get the neutralinos in the mass basis as: 
  %  \begin{equation}
    %  \mathcal{M_D}^0 = N^* \mathcal{M} N^\dagger
    %  \label{eq:diag}
%    \end{equation}
 At tree level, the neutralino-fermion-sfermion couplings can be written as: 
 \begin{eqnarray}
 \mathcal{L}_{\tilde\chi_{\alpha}^0\,\tilde f f} &=& -\sqrt{2}g\overline{\tilde\chi^0_{\alpha}}\left[T^f_3\,N^*_{\alpha\,2}+\tan\theta_W\left(Q_f-T^f_3\right)P_L\tilde f^*_L-\tan\theta_W\, Q_f\,N_{\alpha 1}P_R\tilde{f ^*_R}\right]f \nonumber \\ 
 &-& \frac{g m_f}{\sqrt{2}M_W f_f(\beta)}\overline{\tilde\chi^0_ {\alpha}}\left[N^*_{\alpha h_f} P_L\tilde{f^*_R}N_{\alpha\,h_f}P_R\tilde{f^*_L}\right]f  +  h.c.  
 \label{eq:lagrang}
 \end{eqnarray}
 In the above equation, $N$ denotes the neutralino mass diagonalization matrix, $g$ is the $SU(2)$ gauge coupling, $\theta_W$ the weak mixing angle, $T^f_3$ and $Q_f$ are the third component of the weak isospin and the electric charge of fermion $f$, respectively, $P_{L,R} = (1\pm\gamma_5)/2$ are the chiral projectors, $m_f$ and $M_W$ are the masses of fermion $f$ and of the $W$ boson, respectively, and $\tan\beta$ is the ratio of vacuum expectation values of the two neutral Higgs fields required in the MSSM. Finally, the function $f_f(\beta)$ appearing in eq.(\ref{eq:lagrang}) is $\sin\beta$ for up-type quarks and sneutrinos, and $\cos\beta$ for down-type quarks and charged leptons; similarly the index $h_f = 4\, (3)$ for up-type quarks (down type quarks and charged leptons).
 
 At one loop level eq.(\ref{eq:lagrang}) receives counter-terms and one can extract a flavor independent set of corrections to the neutralino mixing matrix as follows:
    \begin{eqnarray}
    \Delta N_{\alpha 1} &\equiv&  N_{\alpha 1}\left(\frac{\delta g}{g} + \frac{\delta Z_R^{\alpha\alpha}}{2}+\frac{\delta t_W}{t_W}\right) + \sum_{\beta\neq\alpha}{N_{\beta 1}\,\delta Z_R^{\beta\alpha}}\,,\nonumber\\
   \Delta N_{\alpha 2} &\equiv&  N_{\alpha 2}\left(\frac{\delta g}{g} + \frac{\delta Z_R^{\alpha\alpha}}{2}\right) +\sum_{\beta\neq\alpha}{N_{\beta 2}\,\delta Z_R^{\beta\alpha}}\,,\,\nonumber\\
    \Delta N_{\alpha 3} &\equiv&  N_{\alpha 3}\left(\frac{\delta g}{g} + \frac{\delta Z_R^{\alpha\alpha}}{2}+\frac{1}{2}\frac{\delta M_W^2}{M_W^2}-\frac{\delta \cos\beta}{\cos\beta}\right) + \sum_{\beta\neq\alpha}{N_{\beta 3}\,\delta Z_R^{\beta\alpha}}\,\,,\nonumber\\
    \Delta N_{\alpha 4} &\equiv&  N_{\alpha 4}\left(\frac{\delta g}{g} + \frac{\delta Z_R^{\alpha\alpha}}{2}+\frac{1}{2}\frac{\delta M_W^2}{M_W^2}-\frac{\delta \sin\beta}{\sin\beta}\right) + \sum_{\beta\neq\alpha}{N_{\beta 4}\,\delta Z_R^{\beta\alpha}}. 
     \label{eq:Counterterms}
  \end{eqnarray}
Here, $\delta g, \delta t_W, \delta M_W^2, \delta \cos\beta, \delta \sin\beta$ represent the counter-terms of the quantities introduced in eq.(\ref{eq:lagrang}) and $\delta Z_R^{\alpha\beta}$ represent the neutralino wave-function renormalization constants.

Note that eqs.(\ref{eq:Counterterms}) include only two point correlation functions and hence do not include any vertex or box corrections. Also, it should be remembered that these corrections modify mixing matrix $N$ appearing only in the interaction lagrangian.

Eqs.(\ref{eq:Counterterms}) are finite only for the fermion-sfermion loops. At this point a couple of remarks are in order. First, the above expressions are process independent in the sense that they do not depend on the details of external state, second, they were originally introduced to absorb the one loop corrections in the neutralino-fermion-sfermion vertex, however, they can also be used to (partly) absorb all the vertices involving the gaugino component of the neutralino, except for the vertices involving gauge bosons. The most interesting feature of the above set of effective couplings is  the non-decoupling behavior of heavy sparticles. %In case of unbroken suspersymmetry, the fermion-fermion-gauge boson (gauge) coupling $g$ is the same as the corresponding sfermion-fermion-gaugino gauging coupling $\tilde g$. Since supersymmetry is broken at high scale, the heavier sfermions and fermions will decouple from the running of $\tilde g$ below the sfermion mass scale $m_{\tilde f}$. However, the corresponding fermions will still contribute to the running of $g$. Therefore, $g$ and $\tilde g$ run apart below the highest supersymmetry breaking scale. At one loop level we will thus have for scales $Q \leq m_{\tilde f}$:
Due to broken supersymmetry, otherwise equal sfermion-fermion-gaugino coupling $\tilde g$ runs apart from the corresponding fermion-fermion-gauge boson gauge coupling $g$ 
below the sfermion mass scale. At one loop level, this difference is given by:
\begin{equation}
\frac{\tilde g(Q)}{g(Q)} - 1 = \frac{\tilde g(m_{\tilde f})}{g(Q)} - 1 = \beta_h \log\frac{m_{\tilde f}}{Q}, \qquad m_f \leq Q \leq m_{\tilde f}, 
\end{equation}
Thus, correction to $\tilde g$ logarithmically increases with the sfermion mass scale. 

\section{Results}
%\begin{table}[t!]  
 %   \begin{center}
    %  \begin{tabular}{| c | c | c | c | c | c | c | c | c | c | c | c | c | c | c | c | }
%	%\hline Parameter & Value & Parameter & Value\\ 
%	%\hline Parameter & $M_1$ & $Mu_2$ & $M_2$ & $Mu_3$ & $M_3$ & $Md_2$ & $Ml_2$ & $Md_3$ & $Ml_3$ & $A_f$ & $Mr_2$ & $MH_3$ & $Mq_2$ & $\tan\beta$ &  $\mu$\\ 
%	%\hline Value & $90$ & $800$ & $200$ & $800$ & $800$ & $800$ & $250$ & $800$ & $250$ & $0$ & $110$ & $500$ & $800$ & $ 5 $ & $-600$\\ 
%	\hline Parameter & $M_2$ & $Mu_2$ & $M_2$ & $Mu_3$ & $M_3$ & $Md_2$ & $Ml_2$ & $Md_3$ & $Ml_3$\\ 
%	\hline Value & $400$ & $800$ & $200$ & $800$ & $800$ & $800$ & $250$ & $800$ & $250$\\ 
%	\hline
%	\hline Parameter & $A_f$ & $Mr_2$ & $MH_3$ & $Mq_2$ & $\tan\beta$ &  $\mu$\\ 
%	\hline Value & $0$ & $110$ & $500$ & $800$ & $ 5 $ & $-600$\\ 
    %     \hline
     %  \end{tabular}
     %  \label{tab:effbm}
 %   \end{center}
%\end{table}  
The implementation of these effective couplings is done in the on-shell scheme. We use a variant of the on shell scheme for the neutralino sector, as introduced in Chatterjee {\it et.al.}~\cite{Chatterjee:2011wc} We define the electromagnetic coupling constant, $W$ and $Z$-boson masses as on shell quantities. To calculate the relic density, we modify the couplings in micromegas with the above expressions. Taking the input at electroweak scale, we scan over the bino parameter $M_1$. We set $M_A = 500$ GeV, $\tan\beta = 10$, left sfermion soft masses $M_{\tilde l_R} = M_{\tilde q_L} = 500$ GeV, right sfermion soft masses $M_{\tilde u_R} = M_{\tilde d_R} = M_{\tilde Q_L} = 1500$ GeV, $A_f = 1000$ GeV, $M_2 = 400$ GeV and $\mu = 600$ GeV. 
%\subsection{Results}
\begin{figure}[h!]\centering
   \includegraphics[scale=0.4]{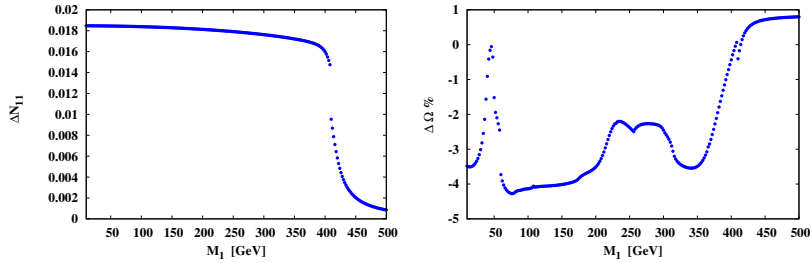}
\caption{ Correction to the bino-like component of the neutralino mixing matrix (left) and the corresponding percentage change to the relic density (right).
\label{fig:effective} }
\end{figure}
In Fig.~\ref{fig:effective} we plot the correction to the bino component of the lightest neutralino $\Delta N_{11}$ (left) and the percent change in the relic density $\Delta\Omega$ (right). Since $M_2 = 400$ GeV, for $M_1 > 400$ GeV, the lightest neutralino is not bino-like anymore. As a result, the correction $\Delta N_{11}$ decreases. While the neutralino is bino-like, the changes in the relic density are characterized by the corrections captured by $\Delta N_{11}$ and opening of various annihilation channels, this is reflected by various peaks appearing in the plot. In some of the annihilation channels only part of the radiative corrections are captured (for example, at $M_1 \approx 84, 94, 106$ GeV, the $W^+ W^-, ZZ, Zh$ channels mediated by a Higgs exchange open and the radiative corrections only to the gaugino-higgisino-neutral higgs vertices are captured.).  
\section{Comparison to full one loop analysis}
This being an effective approach, will fail at some point and it is important to know in which regions of parameter space such couplings are applicable. To this extent we compare the results obtained for $\sigma(\tilde\chi^0_1 \tilde\chi^0_1 \rightarrow \mu^+ \mu^-)$ using effective couplings to those of full one loop results.~\cite{Boudjema:2011ig} The full one loop results were obtained by using SLOOPS.
\begin{figure}[h!]\centering
   \includegraphics[scale=0.5]{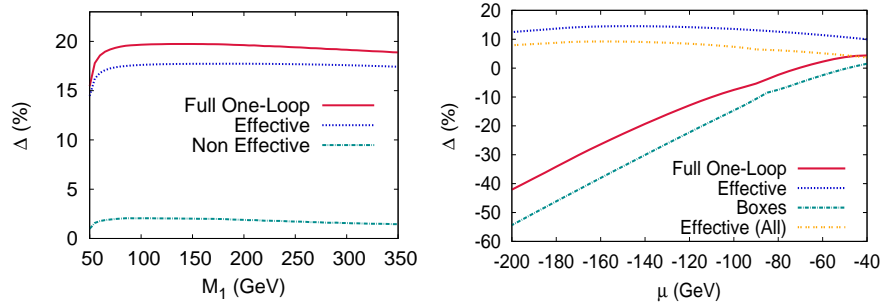}
\caption{Comparison of the full one loop analysis with the effective coupling results, on left, a scan against the bino-like component of neutralino $M_1$ and on right against the higgsinolike component $\mu$. 
\label{fig:compare} }
\end{figure}

In the case, we set $M_A = 1$ TeV, $\tan\beta = 4$, $M_{\tilde l_R} = M_{\tilde l_L} = 500$ GeV, $M_{\tilde u_R} = M_{\tilde d_R} = M_{\tilde Q_L} = 800$ GeV, $A_f = 0$, $M_2 = 500, M_1 = 600$ GeV and $\mu = -600$ GeV. In Fig.~\ref{fig:compare}, left hand side shows the percent change in $\sigma(\tilde\chi^0_1 \tilde\chi^0_1 \rightarrow \mu^+ \mu^-)$ for the full one loop calculations, the effective coupling approach and the difference between the two. We scan over $M_1$ changing the bino fraction of the neutralino. The full one loop analysis shows about $18 ~\textrm{to}~  20\%$ corrections of which about $15 ~\textrm{to}~17\%$ corrections are absorbed by effective couplings remaining non-effective part staying constant at about $2 ~\textrm{to}~3 \%$.~\footnote{Note that we set $\alpha_{em}$ at $\alpha_{em}(q^2=0)$ in this case as opposed to results of Fig.~\ref{fig:effective} where $\alpha_{em}$ is at $\alpha_{em}(q^2=M_Z^2)$ } This is because the neutralino stays bino-like and annihilates dominantly to fermion final states via a sfermion exchange, thus the box and vertex corrections do not contribute significantly. The figure on right shows the percent change in the same cross-section as a function of the higgsino component of the neutralino $\mu$. In this case, the neutralino annihilates via a $Z$ boson exchange and the box and vertex corrections are significant. Effective couplings fail to reproduce the full one loop correction, which ranges from $0 ~\textrm{to}~ -40 \%$. The green line, illustrating the contribution of the boxes is the most dominant contribution, which the effective couplings do not take into account. We tried to implement a new set of effective couplings, displayed in yellow line, however even they fail to reproduce the genuine box diagrams. 

\section{Conclusions}
Effective couplings as used in this work can be an easy and convincing way to include dominant process independent one loop electroweak corrections to neutralino dark matter relic density. These are constructed using two point correlation functions and can be used to absorb corrections to all the gaugino component of the neutralino. When applied to a dominant bino-like neutralino, the changes to the relic density can be as large as $4\%$. They are most effective when the neutralino is bino-like, annihilating into a pair of fermions via sfermion exchange when the contribution from the box and the vertex corrections are negligible. When the neutralino changes it's nature to wino or higgsinolike however, the effective coupling approach fails to reproduce full one loop results. 

\section{Acknowledgements}
I thank the Moriond organizers for financial support to participate in this conference.

\end{document}